\documentclass{aa}
\usepackage{amsmath} 
\usepackage{graphicx}
\usepackage{natbib}
\bibpunct{(}{)}{;}{a}{}{,}
\usepackage{txfonts}
\newcommand{\cmd}{\,cm$^{-2}$}   
\newcommand{\kms}{\,km\,s$^{-1}$}

\newcommand{\myr}{\,$M_{\sun}\,{\rm yr}^{-1}$}
\newcommand{\ha}{H$\alpha$}

\newcommand{\ro}{\,$R_{\sun}$}
\newcommand{\mo}{\,$M_{\sun}$}
\newcommand{\lo}{\,$L_{\sun}$}
\begin{document}

%
%
\title{Wind mass transfer in S-type symbiotic binaries}
\subtitle{II. Indication of  wind focusing}
%
\author{N.~Shagatova, A.~Skopal, \and Z.~Carikov\'a}
\institute{Astronomical Institute, Slovak Academy of Sciences,
        059~60 Tatransk\'{a} Lomnica, Slovakia}
%
\date{Received / Accepted }

\abstract
{
The wind mass transfer from a giant to its white dwarf companion 
in symbiotic binaries is not well understood. For example, the 
efficiency of wind mass transfer of the canonical Bondi-Hoyle accretion mechanism 
is too low to power the typical luminosities of the accretors. 
However, recent observations and modelling indicate a considerably more efficient mass transfer in symbiotic binaries. 
}
{
We determine the velocity profile of the wind from the giant 
at the near-orbital-plane region of eclipsing S-type symbiotic 
binaries EG~And and SY~Mus, and derive the corresponding 
spherical equivalent of the {\rm mass-loss rate. 
With this approach, we indicate the high mass transfer ratio. }
}
{
We achieved this aim by modelling the observed column densities 
taking into account ionization of the wind of  the giant, whose 
velocity profile is derived using the inversion of  Abel's 
integral operator for the hydrogen column density function. 
}
{
Our analysis revealed 
 the spherical equivalent of the 
mass-loss rate from the giant to be  a few times 
$10^{-6}$\myr\ , which is a factor of $\ga 10$ higher than rates 
determined by methods that do not depend on the line of
sight. This discrepancy rules out the usual assumption that the wind is spherically symmetric.
{\rm As our values were derived from  near-orbital-plane 
column densities, these values can be 
a result of focusing the wind from the giant towards 
the orbital plane. }
}
{
Our findings suggests that the wind from giants in S-type  
symbiotic stars is not spherically symmetric, since it is enhanced 
at the orbital plane and, thus, is accreted more effectively 
onto the hot component. 
%
}
\maketitle
\section{Introduction}
\label{int}

Symbiotic stars are binaries with the most distant components 
interacting via a wind mass transfer \citep[e.g.][]{ms99}. 
Usually, the donor is a red giant (RG) and the accretor is a white 
dwarf (WD). The process of accretion heats  the WD up to 
$\gtrsim 10^5$\,K and makes it as luminous as a few times 
$10 - 10^3$\lo. As a result, the hot WD ionizes a large fraction 
of the wind from the giant, giving rise to the symbiotic nebula. 
The so-called S-type systems comprise a normal RG producing
a stellar type of the infrared (IR) continuum, while D-type symbiotics 
contain a Mira-type variable, which produces an additional strong     
emission from the dust in the IR \citep[][]{wa75}.

The wind accretion process represents the principal interaction 
between the binary components; this interaction is responsible for 
the appearance of the symbiotic phenomenon. 
However, already in the 1970s and 80s, 
when the binary nature of these objects was unambiguously  
confirmed by the International Ultraviolet Explorer (IUE), 
an unsolved problem arised: `How do RGs, well within their 
Roche lobes, lose sufficient mass to produce the symbiotic 
phenomenon?' \citep[][]{ken+gall83}. The problem inheres in 
the large measured energetic output from hot components, which 
requires accretion at $10^{-8} - 10^{-7}$\myr, and mass-loss rates of $\approx 10^{-8}$, 
 from M-type giants, that are too low \citep[e.g.][]{reimers77}. 
Values of $\approx 10^{-7}$\myr, as derived for 
RGs in S-type symbiotic stars from radio observations 
\citep[e.g.][]{sea+93,mi02}, do not adequately  explain 
the problem owing to a very low mass transfer efficiency 
(equivalent to the mass transfer ratio, i.e. the mass accretion rate divided by the mass-loss 
rate) of a few percent. Such a low mass transfer efficiency is achieved by a standard 
Bondi-Hoyle wind accretion \citep[][]{b+h44,livwar84}. 
Even more sophisticated 
three-dimensional hydrodynamic simulations showed that the 
mass transfer ratio runs from 0.6 to 10\%, depending on the 
binary configuration \citep[e.g.][]{tbj96,walder97,walder+08}. 
\cite{na+04} found that the mass accretion rate can be even 
smaller than that expected via the simple Bondi-Hoyle approach. 
On the other hand, these authors also recognized a significant increase 
of the mass transfer ratio for very slow winds on the Roche 
surface of the mass-losing star. 

Recent hydrodynamical simulations suggested a very efficient 
wind mass transfer for slow and dense winds, which is typical for 
 asymptotic giant branch stars. In this mass transfer mode, called wind Roche-lobe
overflow \citep[WRLOF;][]{mp07,mp12} or gravitational focusing 
\citep[][]{borro+09}, the wind of the evolved star is focused 
towards the orbital plane and in particular towards the WD, which 
then accretes at a significantly larger rate than from the 
spherically symmetric wind. 
This mode of wind mass transfer was simulated for Mira-type
binaries, and thus can be in effect in D-type symbiotic 
stars. 
An effective wind mass transfer can also be caused by 
a rotation of the mass-losing star. Recently, \cite{sc15} 
applied the wind compression disk model 
\citep[see][]{bjorkcass93,igncassbjork96} to slowly rotating 
normal giants in S-type symbiotic stars, and found that their 
winds can be focused at the equatorial plane with a factor of 
5--10 relative to the spherically symmetric wind. 
Finally, an efficient mass transfer was also evidenced  
observationally for symbiotic stars 
Mira AB \citep[][]{karovska+05}, 
SS Leporis \citep[][]{blind+11,boffin+14a}, 
and FG Serpentis \citep[][]{boffin+14b}. 

In this contribution, we introduce another indication that 
the winds from giants in two eclipsing S-type symbiotic stars, 
EG~And and SY~Mus, can be focused towards the equatorial plane. 
We analyse the measured column densities of neutral hydrogen 
as a function of the orbital phase following the method of \cite{kn93}. 
We improve the application of this method by calculating the ionization 
structure of the wind. 
In Sect.~2, we determine column densities of neutral hydrogen 
around the inferior conjunction of the RG, and in Sect.~3, we 
derive the corresponding theoretical function. The fitting of this  function to 
the observed quantities and resulting parameters are 
introduced in Sect.~4. Discussion of results and conclusion 
can be found in Sects.~5 and 6. 

\section{Column density from Rayleigh scattering}
\label{nHobs}

In the simplest ionization model of a symbiotic binary, 
the neutral region usually has the shape of a cone with 
the giant below its top facing the WD 
\citep[][ hereafter STB]{se84}. 
This implies that the column density of neutral hydrogen on  
the line of sight to the WD, $n^{\rm obs}_{\rm H^0}$, is a function of the orbital phase $\varphi$ and inclination 
$i$ of the binary ($\varphi=0$ at the inferior conjunction 
of the giant). 

The most indicative presence of the neutral hydrogen in 
symbiotic binaries is a strong depression of the continuum 
around the hydrogen lines of the Lyman series caused by 
the Rayleigh scattering \citep[see Fig.~1 of][]{nu89}. 
Assuming it is the dominant process attenuating the WD 
radiation, we can write the observed flux 
$F_{\lambda}(\varphi_j)$ in the $j$-th direction as 
\begin{equation}
 F_\lambda(\varphi_j) = F^0_\lambda
               e^{-n^{\rm obs}_{\rm H^{0}}(\varphi_j)
               \sigma_{\rm Ray}(\lambda)},
\label{rayleighflux}
\end{equation}
where $j = 1, ... , N$, $N$ is the number of $n^{\rm obs}_{\rm H^0}(\varphi_j)$ values at $\varphi_j$, 
$F_{\lambda}^{0}$ is the original (unabsorbed) flux, and 
$\sigma_{\rm Ray}(\lambda)$ is the Rayleigh scattering 
cross section \citep[see Eq.~(5) and Fig.~2 of][]{nu89}. 
For the sake of simplicity, and because of a very high WD 
temperature of $\sim 10^5$\,K, we approximated the far-ultraviolet (far-UV) 
continuum with the black-body radiation. Fitting the function 
(\ref{rayleighflux}) to the observed continuum then yields 
the value of $n^{\rm obs}_{\rm H^0}$ at the given phase. 
This approach was used and described in more detail 
by, for example \cite{inv89}, \cite{vo91}, and \cite{du99}. 

\subsection{Column densities for EG~And and~SY Mus}
\label{nHobsEGaSY}

According to physical and orbital parameters of EG~And and SY~Mus 
\citep[][]{vo92,sc94,hh96,ms99}, their giants are well inside 
the Roche lobes with a corresponding ratio of 
$R_{\rm G}/R_{\rm L} \sim 0.50$. 
Therefore, we can assume that the mass transfer in EG~And
and SY~Mus is proceeding via the wind from their RGs. 
A suggestion that giants in these systems fill around 85\% of 
their Roche lobe is based on the explicit assumption that the 
wave-like variation in their light curves is caused by 
the geometrical effect of the tidal distortion of the surface of the giant \citep[see][]{wv97,ru07}. 

Both systems are eclipsing, showing a deep atmospheric eclipse 
effect caused by the Rayleigh attenuation of the far-UV continuum 
\citep[][]{vo90,pe96,du99,pe99}. \cite{vo92} estimated the inclination 
of the EG~And orbit $i = 78.5^{\circ} - 90^{\circ}$ with 
a confidence of 72\%. Therefore, we  consider that $i > 70^{\circ}$. 
For SY~Mus, \cite{hh96} determined $i = 84.2\pm 1.7^{\circ}$ from 
its spectropolarimetric orbit. 

\begin{figure}
\centering
\begin{center}
\resizebox{\hsize}{!}{\includegraphics[angle=0]{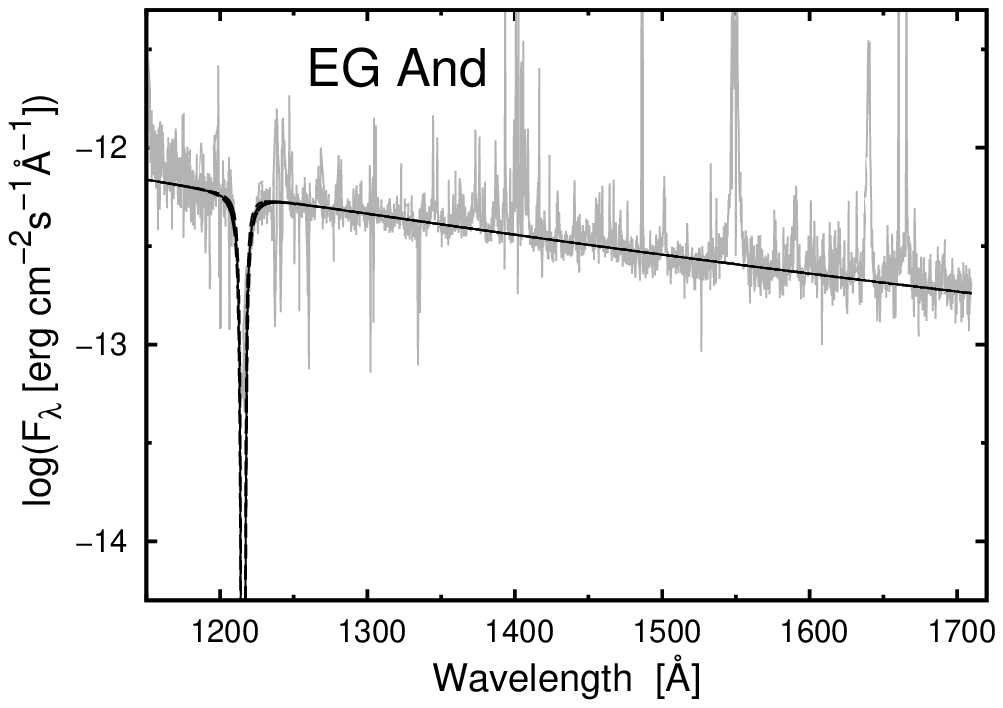}}
\resizebox{\hsize}{!}{\includegraphics[angle=0]{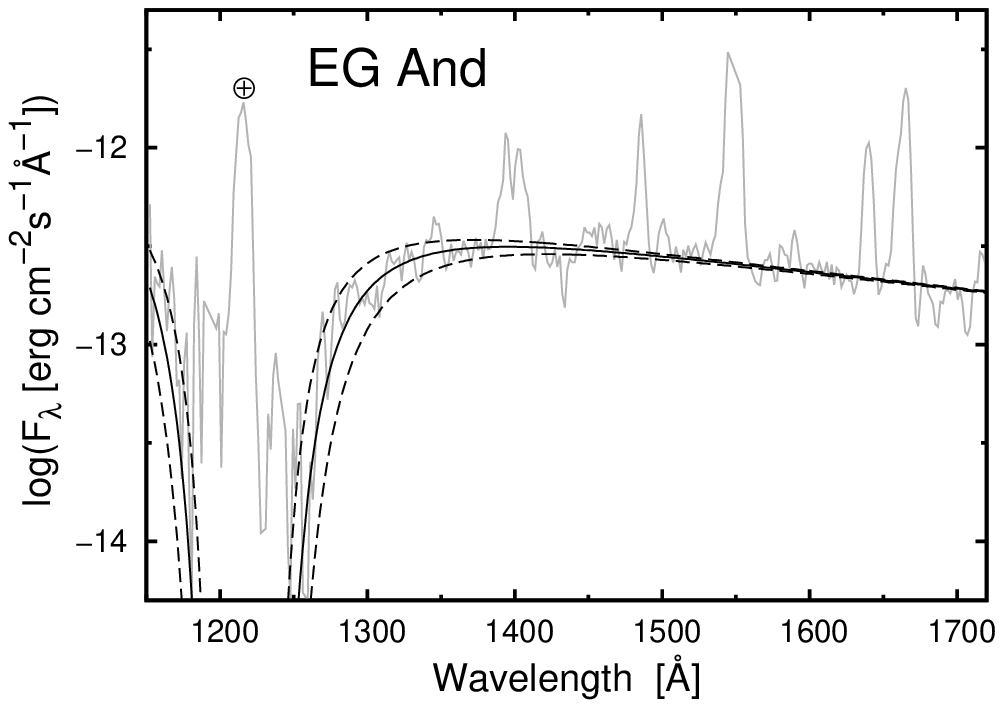}}
\end{center}
\caption[]{
Examples of the far-UV spectrum of EG~And (grey line) compared 
to models (solid and dashed lines) given by 
Eq.~(\ref{rayleighflux}) at different orbital phases. 
Top: 
$\varphi = 0.1695$, 
$n^{\rm obs}_{\rm H^0} = (3\pm 1)\times 10^{20}$\cmd. 
Bottom: 
$\varphi = 0.0847$, 
$n^{\rm obs}_{\rm H^0} = (1.5 +0.7/-0.5)\times 10^{23}$\cmd\ 
(see Table~1). $\oplus$ denotes the geocoronal Ly-$\alpha$ 
emission. 
          }
\label{HST1IUE2}
\end{figure}

\begin{figure}
\centering
\begin{center}
\resizebox{\hsize}{!}{\includegraphics[angle=0]{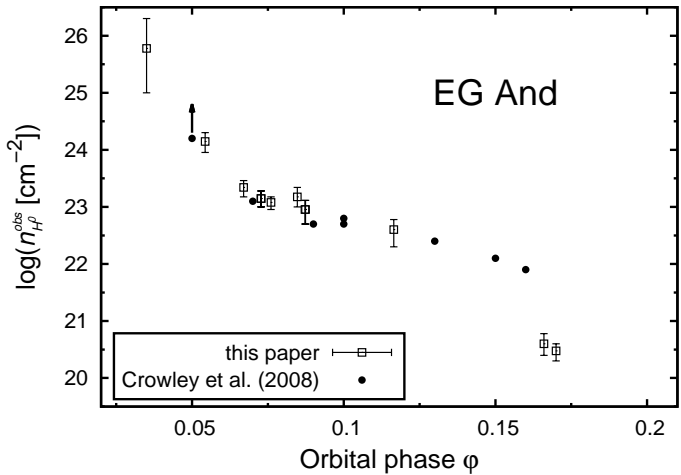}}
\end{center}
\caption[]{
Egress column densities as a function of the orbital phase 
for EG~And. Our values (Table~1) and those of \cite{cr08} 
are depicted by squares and circles, respectively. 
          }
\label{EGAndnH}
\end{figure}

In fitting the observed continuum (1), we approximated the flux 
emitted by the WD with that of a black body radiating at  
temperatures of 95\,000\,K and 110\,000\,K for EG~And 
\citep[][]{sk05} and SY~Mus \citep[][]{mu91}, respectively. 
For EG~And, we determined new values of $n^{\rm obs}_{\rm H^0}$ 
using the 
IUE spectra from 1980 -- 1993 and the Hubble Space Telescope (HST) 
spectra obtained in 2002 and 2009 (Table~1). EG~And spectra were 
dereddened with $E_{\rm B-V} = 0.05$\,mag \citep[][]{mu91} 
and SY~Mus spectra were dereddened with $E_{\rm B-V} = 0.35$\,mag 
\citep[][]{sk05}. 
\begin{table}[t!]
   \caption{
Column densities of neutral hydrogen $n_{\rm H^0}^{\rm obs}$ 
we measured on UV IUE and HST(STIS) spectra of EG~And 
(Sect.~2). 
           }
\label{nHobsT}
\centering
\begin{tabular}{cccc}
\hline\hline
MJD & $\varphi$ & $n_{\rm H^0}^{\rm obs}$ [cm$^{-2}$]
    & Spectrum  \\ 
\hline
52630.37 & 0.0348 & $(6.0+14.0/-5.0)\times 10^{25}$& O8C406040\\  
49261.70 & 0.0543 & $(1.4+0.6/-0.5)\times 10^{24}$ & SWP48820 \\  
55058.75 & 0.0666 & $(2.2+0.7/-0.7)\times 10^{23}$ & OB7503020\\  
49270.60 & 0.0727 & $(1.4+0.5/-0.4)\times 10^{23}$ & SWP48883 \\  
49270.64 & 0.0728 & $(1.4+0.5/-0.4)\times 10^{23}$ & SWP48884 \\  
55062.81 & 0.0750 & $(1.2+0.3/-0.3)\times 10^{23}$ & OB7504020\\  
44450.68 & 0.0847 & $(1.5+0.7/-0.5)\times 10^{23}$ & SWP09644 \\  
49277.58 & 0.0872 & $(9.0+4.0/-4.0)\times 10^{22}$ & SWP48938 \\  
49277.65 & 0.0874 & $(9.0+4.0/-4.0)\times 10^{22}$ & SWP48939 \\  
49291.69 & 0.1165 & $(4.0+2.0/-2.0)\times 10^{22}$ & SWP49056 \\  
55106.42 & 0.1654 & $(4.0+2.0/-1.5)\times 10^{20}$ & OB7505010\\  
55108.42 & 0.1695 & $(3.0+1.0/-1.0)\times 10^{20}$ & OB7506020\\  
\hline
\end{tabular}
\end{table}
We adopted other $n^{\rm obs}_{\rm H^0}$ values from \cite{cr08}. 
The IUE spectra from the ingress part of the UV light curve 
(i.e. obtained prior to the inferior conjunction of the giant) 
were too noisy to be used. Therefore, we only modelled the data 
from the egress, i.e. after the giant conjunction. Quantities of 
$n^{\rm obs}_{\rm H^0}$ for SY~Mus were adopted from \cite{du99}. 
They cover both the ingress and egress part of the light curve. 
We also justified the adopted $n^{\rm obs}_{\rm H^0}$ values for 
both stars and found a good agreement with our results. 
Our estimate of uncertainties was $\approx 30 - 50\%$, which is comparable 
with that of \cite{du99}. 
Figure~\ref{HST1IUE2} shows examples of the Rayleigh attenuation 
models at different orbital phases, while Fig.~\ref{EGAndnH} 
summarizes all $n^{\rm obs}_{\rm H^0}$ values for EG~And. 
For EG~And and SY~Mus, we used the ephemeris for the inferior 
conjunction of the giant according to \cite{fe00} and \cite{du99}, 
respectively. 
%
%
%
%
%
\section{The method}
\label{theory}
\subsection{Theoretical column densities}
\label{invpr}

Throughout this paper, we consider a generally used assumption that the wind is spherically symmetric. However, our results (Table~\ref{tabMdot}) contradict this requirement. Thereby, we use the proof by contradiction to demonstrate the presence of wind focusing (Sect.~\ref{sect:focusing}). 

In the spherically symmmetric model of the wind,  its continuity equation can be written as
%
\begin{equation}
   \dot M=4\pi r^2N_{\rm H}(r)\mu m_{\rm H}v(r),
\label{cont}
\end{equation} 
where $\dot M$ is the mass-loss rate from the giant, $r$ 
the distance from its centre, $N_{\rm H}(r)$ the total hydrogen number density, $\mu$ the mean molecular weight, $m_{\rm H}$ the mass 
of the hydrogen atom, and $v(r)$ the velocity of the wind. 
If we integrate $N_{\rm H}(r)$ given by Eq.~(\ref{cont}) along 
the line-of-sight $l$ from the observer ($-\infty$) to the 
infinity containing the WD, the column density of total 
hydrogen, $n_{\rm H}$, can be expressed as 
\begin{equation}
  n_{\rm H} = \displaystyle\frac{\dot M}{4\pi \mu m_{\rm H}}
               \displaystyle\int\limits_{-\infty}^{\infty}
               \displaystyle\frac{{\rm d}l}{r^2v(r)}. 
\label{nH}
\end{equation}
Using the separation between the binary components $p$, 
projected to the plane perpendicular to the line of sight, the so-called impact parameter $b$ \citep[e.g.][]{vo91}, 
\begin{equation}
   b^2=p^2(\cos^2{i}+\sin^2\varphi\sin^2i),
\label{b2}
\end{equation}
and the relation $r^2=l^2+b^2$ \citep[e.g. Fig.\,1 of][]{kn93},
we can rewrite Eq.~\eqref{nH} to the form
\begin{equation}
 \displaystyle n_{\rm H}(b) = a\int\limits_{b}^{\infty}
  \displaystyle\frac{{\rm d}r}{rv(r)\sqrt{r^2-b^2}},
\label{nHab}
\end{equation}
where 
\begin{equation}
  a = \displaystyle\frac{2\dot{M}}{4\pi\mu m_{\rm H}}.
\label{aexp}
\end{equation}
If the wind velocity profile (WVP), $v(r)$, is known, we can 
directly calculate $n_{\rm H}$ at a given phase, since the 
distances $l$ and $r$ depend on $\varphi$. However, the 
situation is just the opposite; we know $n_{\rm H}$ from 
observations, but the WVP is unknown. So, we have an inversion 
problem, which was solved by \cite{kn93}.  


\subsection{The wind velocity profile formula}
\label{wlexpr}

Basically, the method of \cite{kn93} allows us to determine 
the WVP with the inversion of the integral operator for column 
density by its diagonalization. Since this inversion problem 
is generally unstable, the diagonalization is restricted to 
the finite series with an appropriate number of terms. 
According to \cite{du99}, the adequate finite sum representing 
the column density in the inversion method can be expressed as 

\begin{equation}
\label{nHr}
   \tilde{n}_{\rm H}(b) = \displaystyle\frac{n_1}{b} + 
                  \displaystyle\frac{n_K}{b^K},
\end{equation}
where coefficients $n_1$, $n_K$, and $K$ can be obtained by fitting 
the observed $n^{\rm obs}_{\rm H^0}$ values (see Sec.~\ref{proc}). 
The first term dominates for large values of $b$, while the second 
term dominates for small values of $b$. It is not necessary to use the 
$\tilde{n}_{\rm H}(b)$ function with more terms because of the large 
errors in the $n^{\rm obs}_{\rm H^0}$ data (Tab.~\ref{nHobsT}). 
%
Moreover, Eq.~(\ref{nHr}) represents a good approximation of 
the original $n_{\rm H}(b)$ function (\ref{nHab}) 
and has a simple inverted form, 
\begin{equation}
  \displaystyle\frac{a}{rv(r)} = 
                \displaystyle\frac{n_1}{\lambda_1r} + 
                \displaystyle\frac{n_K}{\lambda_Kr^K} .
\label{eq:gr}
\end{equation}
For $r\rightarrow\infty$, 
$\displaystyle\frac{a}{rv(r)} \rightarrow 
\displaystyle\frac{n_1}{\lambda_1r}$ 
and $v(r)\rightarrow v_{\infty}$, where $v_\infty$ is 
the terminal velocity of the wind, we obtain the relation 
\begin{equation}
   \displaystyle\frac{a}{v_\infty} = 
   \displaystyle\frac{n_1}{\lambda_1},
\label{ainf}
\end{equation}
and finally, using Eq.~(\ref{eq:gr}), the WVP has the form 
\begin{equation}
\label{wind}
   v(r) = \displaystyle\frac{v_\infty}{1 + \xi r^{1-K}},
\end{equation}
where $\xi = \displaystyle\frac{n_K\lambda_1}{n_1\lambda_K}$, 
$\lambda_1$, and $\lambda_K$ are the eigenvalues of the Abel 
operator given by the relation \citep{kn93}
\begin{center}
  $\hspace{-1.4cm}i=1
   \hspace{1.9cm} \lambda_1 = \displaystyle\frac{\pi}{2}$,
\end{center}
\begin{equation}
\label{vlhodn}
  i\geq 2 \hspace{2cm} 
          \lambda_i = \displaystyle\frac{\pi}{2(i-1)\lambda_{i-1}}.
\end{equation}
Equation~(\ref{wind}) is sufficiently flexible to represent a great variety of velocity profiles (e.g. to be comparable to the widely used $\beta$-law; see Fig.~\ref{windXiK}). 

To summarize, the parameters $n_1$, $n_K$, and $K$ can be obtained 
by fitting the measured $n_{\rm H^0}^{\rm obs}$ data with 
Eq.~(\ref{nHr}), and thus the WVP in the form of Eq.~(\ref{wind}). 
However, since the giant wind is partially ionized, we consider 
this effect in our approach. 
%

\subsection{Effect of the ionization on $\rm H^0$ column density}
\label{ionstruct}

During quiescent phases, a fraction of the neutral wind from 
the giant is ionized by the hard radiation emitted by the hot 
component. Hence, the shape of the neutral $\rm H^0$ region 
is determined by the $\rm H^0$/$\rm H^+$ boundary. Therefore, 
to calculate a value of $\rm H^0$ column density, we have to 
integrate Eq.~(\ref{nH}) from the observer to the intersection 
of the line of sight with the $\rm H^0$/$\rm H^+$ boundary, 
$l(\varphi)$ (see Fig.~2 of \cite{ss12}), i.e. 
\begin{equation}
  \displaystyle n_{\rm H^0} = \frac{a}{2}
  \displaystyle\int\limits_{-\infty}^{l_{\varphi}}
  \displaystyle\frac{dl}{r^2v(r)} .
\label{nHH}
\end{equation} 
Using the parameter $b$ (Eq.~(\ref{b2})), which determines the 
direction of the line-of-sight $l$ to the WD, Eq.~(\ref{nHH}) 
can be rewritten as 
\begin{equation}
  n_{\rm H^0}(b) = \displaystyle\frac{a}{2}
          \displaystyle\int\limits_{-\infty}^{l_{\varphi}(b)}
          \displaystyle\frac{dl}{(l^2+b^2)v(\sqrt{l^2+b^2})},
\label{nHnum}
\end{equation}
which we use to model the $n_{\rm H^0}^{\rm obs}$ data 
(Sect.~\ref{mod}). 

Finally, the $\rm H^0$/$\rm H^+$ interface can be obtained 
from the condition of the ionization/recombination equilibrium 
between the flux of ionizing photons from the WD and  rate 
of recombinations in the wind from the giant. 
According to \cite{nv87}, the solution of the parametric equation 
\begin{equation}
   f(u,\vartheta) - X^{\rm H+} = 0,
\label{XH+=f}
\end{equation}
defines the boundary between neutral and ionized gas at the 
orbital plane determined by a system of polar coordinates 
($u,\vartheta$) with the origin at the hot star. 
The ionization parameter $X^{\rm H+}$ is written as
\begin{equation}
   X^{\rm H+} = \frac{4\pi(\mu m_{\rm H})^2}
                 {\alpha_{\rm B}({\rm H},T_{\rm e})} p L_{\rm ph}
            \left(\frac{v_\infty}{\dot M}\right)^2, 
\label{XH+}
\end{equation}
where $L_{\rm ph}$ is the flux of ionizing photons from the WD 
and $\alpha_{\rm B}({\rm H},T_{\rm e})$ is the total hydrogen 
recombination coefficient for recombinations other than to the 
ground state (Case B) at the electron temperature, $T_{\rm e}$. 
The function $f(u,\vartheta) \propto v(r)^{-2}$ was treated 
for the first time by STB for a steady state situation 
and a constant wind velocity. 
Examples of the $\rm H^0$/$\rm H^+$ boundaries for different 
WVP can be found in, for example \cite{nv87}, \cite{sk01}, and 
\cite{ss12}. Intersection of the line of sight with the boundary 
is determined by the parameter $l_{\varphi}(b)$ 
\citep[see Sect.~3.3 of][]{ss12}. 

As a result of a very rapid transition of wind hydrogen from 
an almost completely neutral to an almost fully ionized state 
at the ionization boundary, the flux of ionizing photons penetrating 
 the neutral region is negligible and the flux of the neutral 
hydrogen atoms to the ionized zone is even of less importance 
(see e.g. Figs.~2c and 6 of \cite{sc97}, or Fig.~4.14 of \cite{cr06}).
%
%
%
\begin{table*}[t!]
\caption{Resulting parameters ($n_1$, $n_{\rm K}$, $K$ and 
         $X^{\rm H+}$) of modelling $n_{\rm H^0}^{\rm obs}$ 
         column densities (Sect.~\ref{nHobsEGaSY}) with the 
         function (\ref{nHnum}). 
         They define the WVP in the form of Eq.~(\ref{wind}) 
         and the ionization structure of the wind given by 
         the solution of Eq.~(\ref{XH+=f}). 
}
\label{vysl}
\centering
\begin{tabular}{lccccccccc}
\hline\hline
object &$i$ &E/I$^{1)}$ & $n_1[10^{23}]$  & $n_{\rm K}$ & 
  $K$ & $\xi$ & $X^{\rm H+}$& $\chi_{\rm red}^2$ & model \\
    \hline
EG~And
& $70^{\circ}$ & E & $4.54$ & $5.12\times 10^{30}$ & $21$ & 
  $6.40\times 10^{7}$ & $1.75$&1.67 & I \\
& $80^{\circ}$ & E & $3.87$ & $1.15\times 10^{27}$ & $14$ & 
  $1.38\times 10^{4}$ & $1.85$&1.60 & J \\
& $90^{\circ}$ & E & $3.40$ & $4.83\times 10^{25}$ & $10$ & 
  $5.49\times 10^{2}$ & $1.88$&1.63 & K \\
SY~Mus 
& $80^{\circ}$ & E & $8.00$ & $1.50\times 10^{27}$ & $9 $ &
  $8.94\times 10^{3}$ & $2.30$ & 1.02 & L \\
& $84^{\circ}$ & E & $6.20$ & $5.30\times 10^{26}$ & $8 $ &
  $2.94\times 10^{3}$ & $2.50$ & 0.94 & M \\
& $90^{\circ}$ & E & $6.10$ & $5.00\times 10^{26}$ & $8 $ & 
  $2.82\times 10^{3}$ & $2.53$ & 1.39 & N \\
& $84^{\circ}$ & I & $2.45$ & $1.00\times 10^{27}$ & $13$ & 
  $1.81\times 10^{4}$ & $16.0$ & 2.33 & O \\
\hline
\multicolumn{10}{l}{Notes: $^{1)}$ E -- egress data, I -- ingress data}
\end{tabular}
\end{table*}

\section{Modelling and results}
\label{mod}

\subsection{The procedure}
\label{proc}
In this section, we determined the WVP and the corresponding 
$\dot M$ for giants in EG~And and SY~Mus by fitting the measured 
column densities $n_{\rm H^0}^{\rm obs}$ with the function 
(\ref{nHnum}). This indicates we need to find parameters $n_1$, $n_K$ and $K$, 
which determine the WVP in the form of Eq.~(\ref{wind}) and  
parameter $X^{\rm H+}$, which determines the geometry of the 
$\rm H^0$/$\rm H^+$ boundary. To avoid unreasonably high values 
of $n_1$ and $n_K$, we used the giant radius, $R_{\rm G}$, as 
the length unit.
First, we estimated possible ranges of fitting parameters with 
a preliminary match of the data with the function (\ref{nHnum}). 
Then, we proceeded as follows: 

(i)
Using the Monte Carlo method, we assigned values of all 
searched parameters, $n_1$, $n_K$, $K$, and $X^{\rm H+}$, 
within the given ranges. 

(ii)
For the corresponding WVP and the $X^{\rm H+}$ 
parameter, we solved Eq.~(\ref{XH+=f}) to get the intersection 
of the line of sight with the ionization boundary determined 
by the parameter $l_{\varphi}(b)$. 

(iii)
Then, we calculated the $n_{\rm H^0}(b)$ function 
(Eq.~(\ref{nHnum})) for the model WVP and $l_{\varphi}(b)$, 
compared this function with the $n_{\rm H^0}^{\rm obs}$ data, and obtained 
the reduced $\chi$-squared sum, $\chi^2_{\rm red}$. 

(iv)
Finally, iterating this procedure within the space of input 
parameters provided us a set of models. We selected the best solution 
 using the condition of the least-squares method. 

In the case of SY~Mus, the best fits were found empirically 
because of scanty data for $b<2.5$ (see Fig.~\ref{nHSYMus}). 
To match the ingress data, we used relation 
(see Eq.~(\ref{XH+}))
\begin{equation}
X^{\rm H+} \propto \displaystyle\left(\dfrac{v_\infty}
                                      {\dot{M}}\right)^2,
\label{Xsim}  
\end{equation}
to obtain the ingress/egress $n_1$ parameters ratio 
(see Eq.~(\ref{eq:dotm}) below), and thus the ingress $n_1$ 
value for the model egress $n_1$ parameter. 

\subsection{The spherical equivalent of the mass-loss rate}
\label{sect:Msp}

The inversion method allows us
to determine 
the value of the corresponding mass-loss rate in the spherically symmetric model, i.e. the spherical equivalent of the mass-loss rate $ \dot{M}_{\rm sp}$. 
By analogy to Eq.~(\ref{ainf}) for the integral (\ref{nHab}), 
we can write 
\begin{equation}
  \dfrac{a}{2R_{\rm G}} = \dfrac{n_1v_\infty}{\lambda_1}
\end{equation}
for the integral (\ref{nHnum}), expressed in the length units of 
$R_{\rm G}$, for $r\rightarrow \infty$. 
Then, using Eq.~(\ref{aexp}) we obtain the relationship 
\begin{equation}
 \dot{M}_{\rm sp} = 4\pi\mu m_{\rm H}R_{\rm G}
                           \dfrac{n_1}{\lambda_1}v_\infty ,
\label{eq:dotm}
\end{equation}
where $n_1$ is the resulting parameter of our models 
(see Table~\ref{vysl}). 

Comparison of $\dot{M}_{\rm sp}$ quantities (Table~3) with 
the total mass-loss rates, $\dot M$, determined by methods that 
do not depend on the line of sight (Sect.~5.1), allows us to 
test whether the wind from giants in S-type symbiotic binaries 
is spherically symmetric or not (Sect.~\ref{sect:focusing}). 

\subsection{Estimate of uncertainties}
\label{uncert}

The primary source of uncertainties of the fitting parameters 
is given by the errors of the measured column densities 
$n_{\rm H^0}^{\rm obs}$ (Table~1). 
Therefore, to estimate possible ranges of the model parameters, 
$n_1$, $n_K$, $K$, and $X^{\rm H+}$, we prepared a grid of 
corresponding models (\ref{nHnum}) around the best solution to cover the measured column densities within their errors. 
In this way, we estimated uncertainties in $n_1$ to be in 
the range of 10--30\%, which, according to Eq.~(\ref{eq:dotm}), 
can also be assigned to $\dot M$. 
We found small uncertainties around 
of 3--6\% for the parameter $X^{\rm H+}$ because 
its value is given by the orbital phase, at which the line of 
sight coincides with the tangent to the $\rm H^0$/$\rm H^+$ 
boundary (i.e. when $n_{\rm H^0}^{\rm obs} \rightarrow 0$). 
In both cases, this tangent is well defined by the smallest values of 
$n_{\rm H^0}^{\rm obs}$ at $\varphi \sim$~0.17 and 0.16 for 
EG~And and SY~Mus. 
The $K$ index is always within 10\% of its most probable 
values. 
However, the parameter $n_{\rm K}$ is very uncertain ($> 70\%$) 
because its value dominates the $\tilde n_{\rm H}(b)$ function 
(Eq.~(\ref{nHr})) for small $b$, where the $n_{\rm H^0}^{\rm obs}$ 
measurements suffer from large errors (Table~1). 

A further source of uncertainties is given by the assumption of the spherical symmetry. Since the H$^0$ column density data show the ingress/egress asymmetry and the main result of the wind focusing (see below Sect.~\ref{sect:focusing}) rules out the spherically symmetric wind, the WVPs derived in this paper are also affected by the following systematic error. However, it is hard to assess its magnitude.
Since there is a lack of observations resolving spatial wind distribution for S-type symbiotic systems, our approach can provide, at least, some constraints on theoretical modelling of the circumstellar matter in symbiotic binaries.

Another source of the systematic error is given by using 
a subset of the full space of eigenfunctions of Abel operator 
to avoid ill-conditioning, as proposed by \cite{kn93}. Since 
the column density models represent data well, we suppose 
this uncertainty to be of less magnitude than errors from 
other mentioned sources.

\subsection{Application to EG~And and SY~Mus}

\begin{figure}
\centering
\begin{center}
\resizebox{\hsize}{!}{\includegraphics[angle=0]{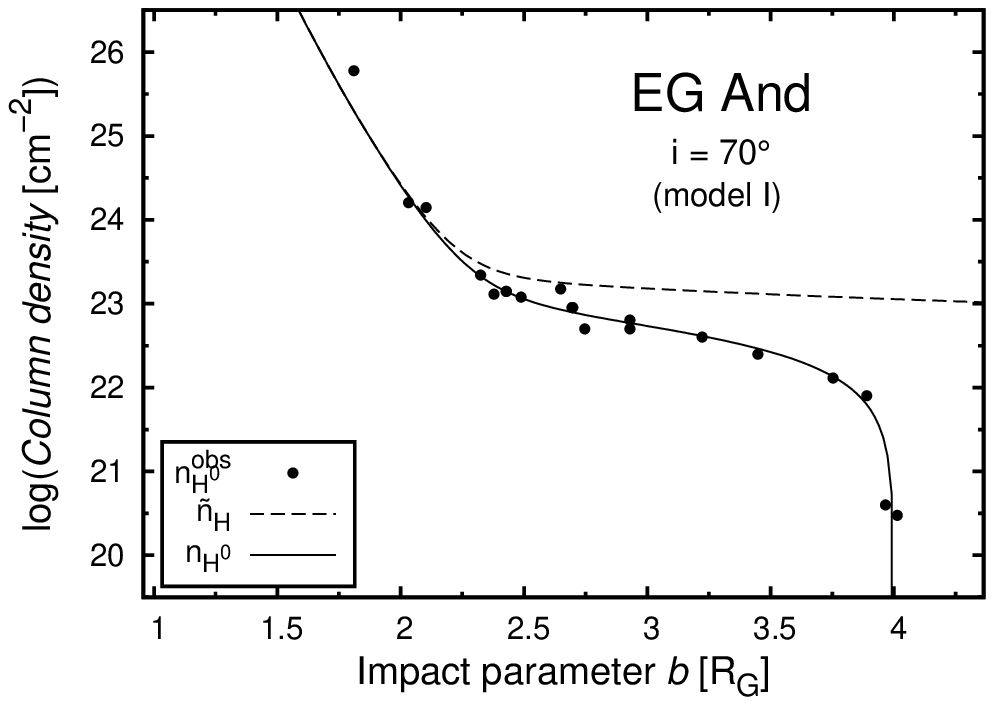}}
\resizebox{\hsize}{!}{\includegraphics[angle=0]{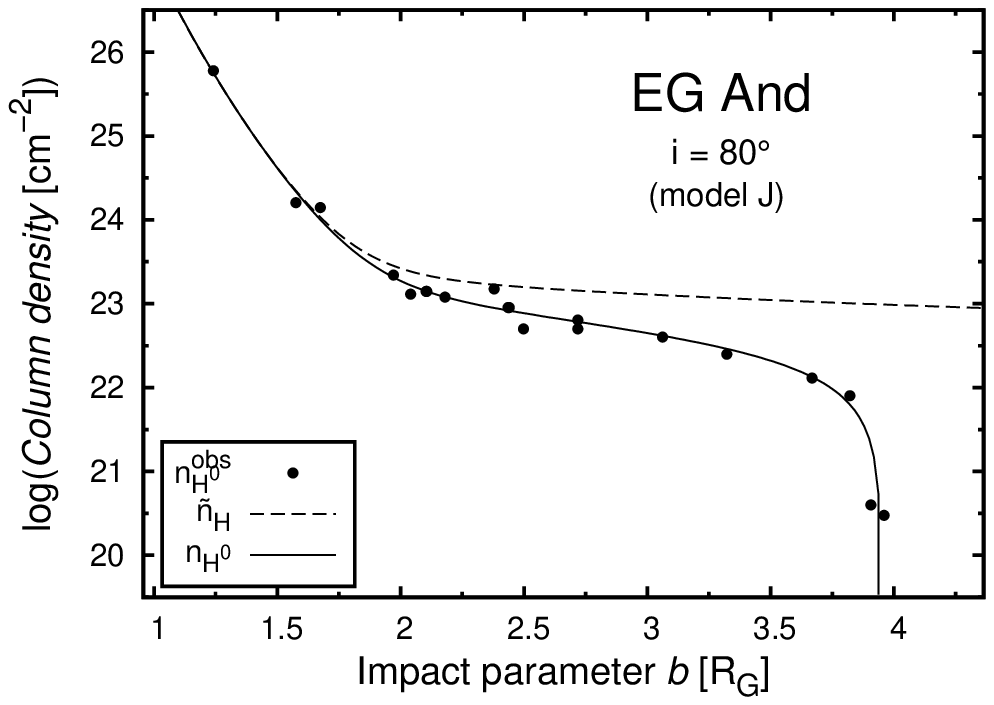}}
\resizebox{\hsize}{!}{\includegraphics[angle=0]{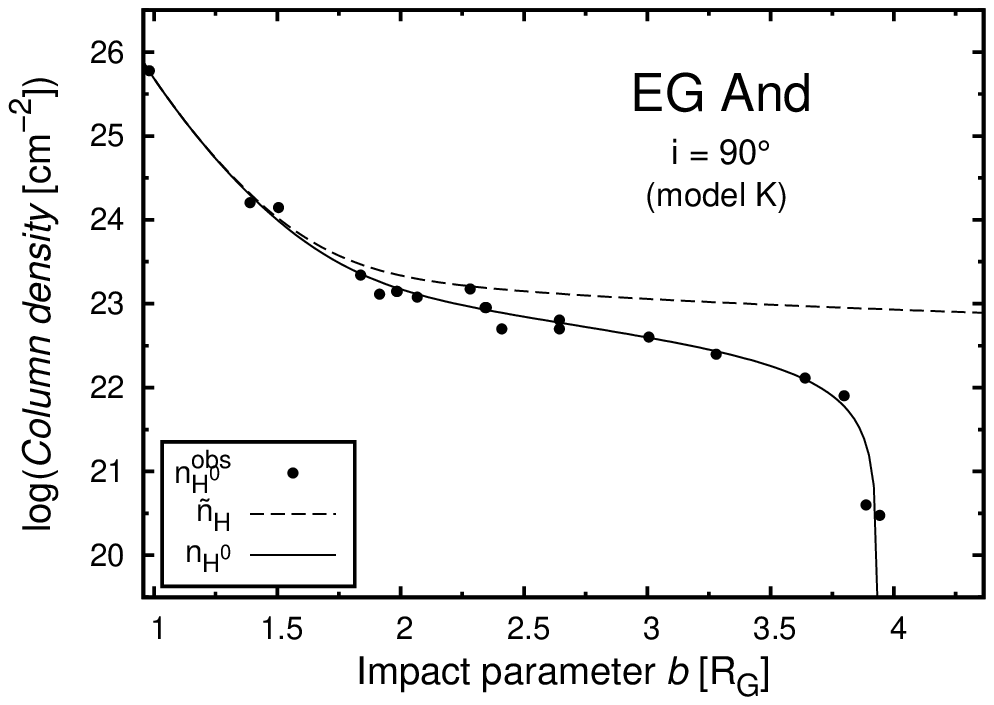}}
\end{center}
\caption[]{
Resulting fits of $n_{\rm H^0}^{\rm obs}(b_j)$ data for EG~And 
(dots) with the $n_{\rm H^0}(b)$ function (Eq.~(\ref{nHnum}); 
solid lines). Parameters of models are listed in Table~\ref{vysl}. 
Corresponding $\tilde{n}_{\rm H}(b)$ functions, which represent 
the total hydrogen column densities (i.e. no ionization included, 
Eq.~(\ref{nHr})) are depicted with dashed lines. 
          }
\label{nHEgressEGAnd}
\end{figure}

\begin{figure}
\centering
\begin{center}
\resizebox{\hsize}{!}{\includegraphics[angle=0]{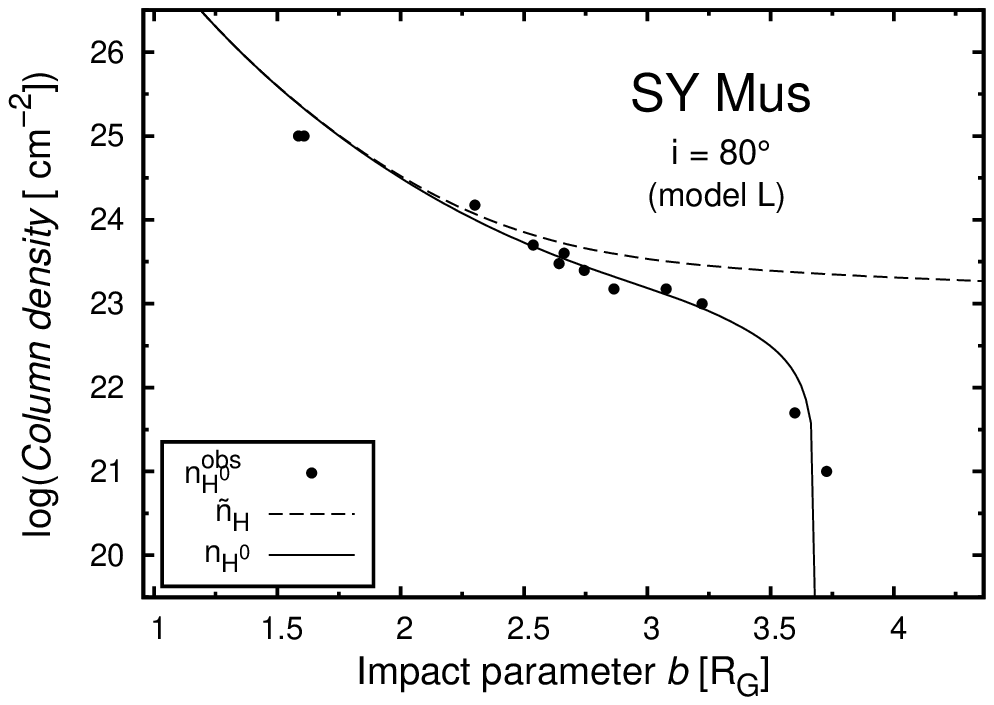}}
\resizebox{\hsize}{!}{\includegraphics[angle=0]{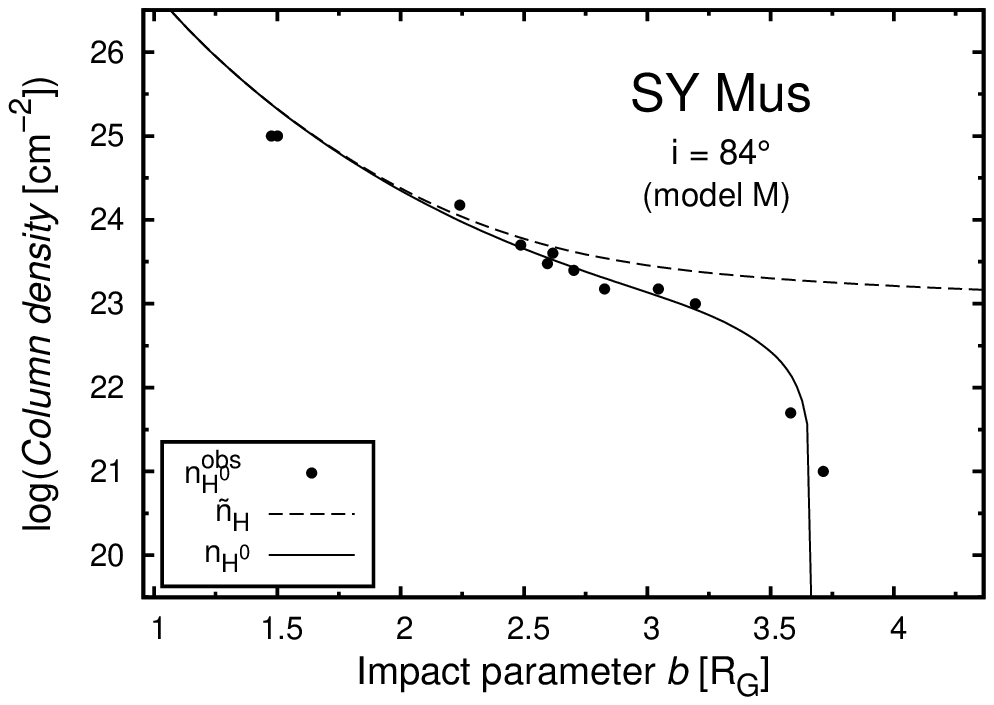}}
\resizebox{\hsize}{!}{\includegraphics[angle=0]{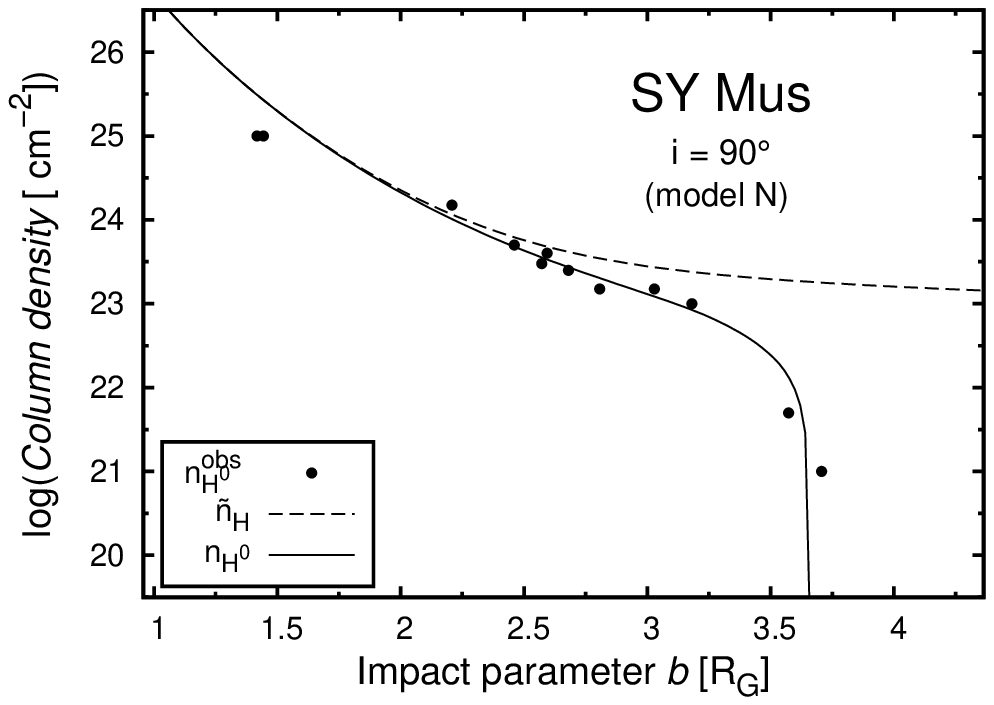}}
\resizebox{\hsize}{!}{\includegraphics[angle=0]{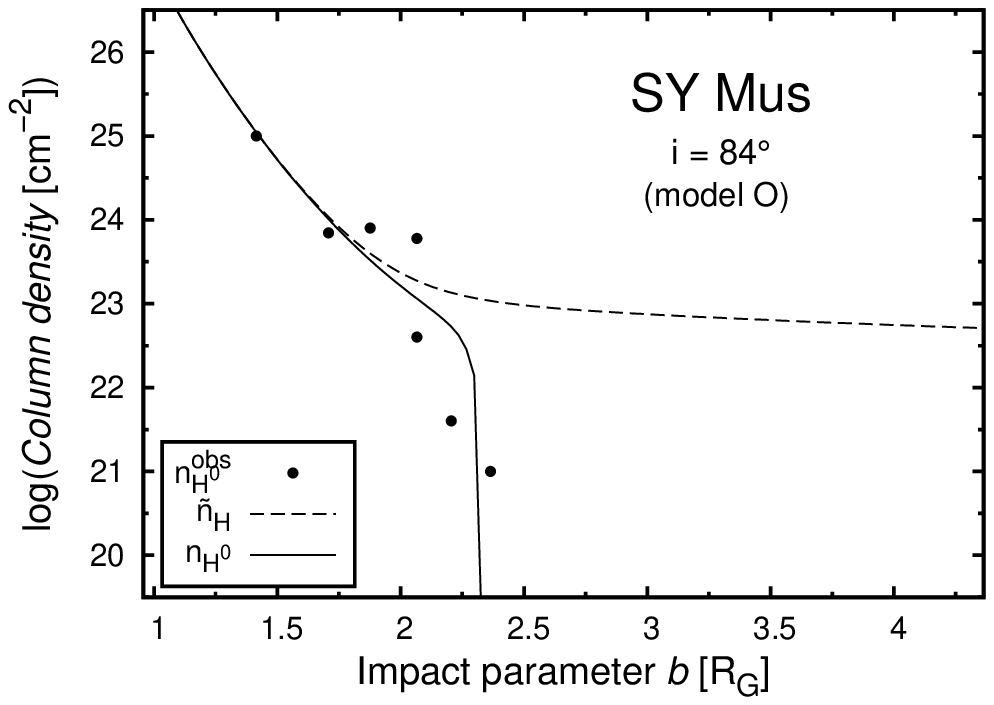}}
%
\end{center}
\caption[]{
As in Fig.~\ref{nHEgressEGAnd}, but for SY~Mus data. 
          }
\label{nHSYMus}
\end{figure}

According to the geometry of the $\rm H^0$ zone in the binary, 
as given by $n_{\rm H^0}^{\rm obs}$ measurements 
(e.g. Fig.~\ref{EGAndnH}), we calculated the 
$n_{\rm H^0}(b)$ function (\ref{nHnum}) for the range of orbital 
phases $\left\langle 0.02, 0.21\right\rangle$ and adopted 
orbital inclinations $i$. 
According to \cite{vo92} and \cite{hh96}, $i > 70{\degr}$, 
$p = 330$\ro\ and $i = 84{\degr}.2\pm 1{\degr}.7$, $p = 376$\ro\ 
for EG~And and SY~Mus, respectively. 
Therefore, we selected $i = 70, 80, 90{\degr}$ 
and $i = 80, 84, 90{\degr}$ in fitting the 
$n_{\rm H^0}^{\rm obs}$ values of EG~And and SY~Mus. 
Resulting parameters are summarized in Table~\ref{vysl} 
and the corresponding models are shown in 
Figs.~\ref{nHEgressEGAnd} and \ref{nHSYMus}. 
For the EG~And model with $i = 90{\degr}$, the highest value of 
$n_{\rm H^0}^{\rm obs}$ has $b\lesssim1$, which suggests that 
$i < 90{\degr}$ because the line of sight would pass through 
the giant itself. Nevertheless, we show this solution just for 
comparison. 

To determine 
the spherical equivalent of the mass-loss rate from 
the giant according to Eq.~(\ref{eq:dotm}), we used 
$R_{\rm G}$ = 75 and 86\ro\ for EG~And \citep[][]{vo92} and 
SY~Mus \citep[][]{sc94} and characteristic 
$v_\infty$ = 20--50\kms\ (Sect.~\ref{vnekdotM}). 
Examples for $v_\infty$ = 30\kms\ are introduced in 
Table~\ref{tabMdot}. 
%
%
%
%
%
%
\begin{figure}
\centering
\begin{center}
\resizebox{\hsize}{!}{\includegraphics[angle=0]{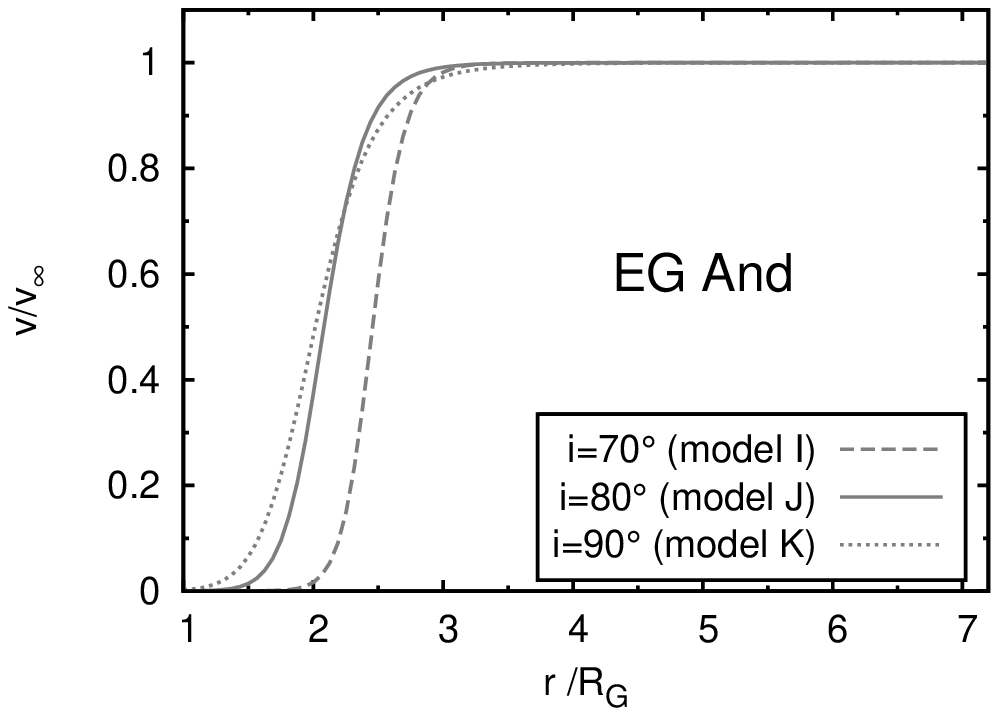}}
\resizebox{\hsize}{!}{\includegraphics[angle=0]{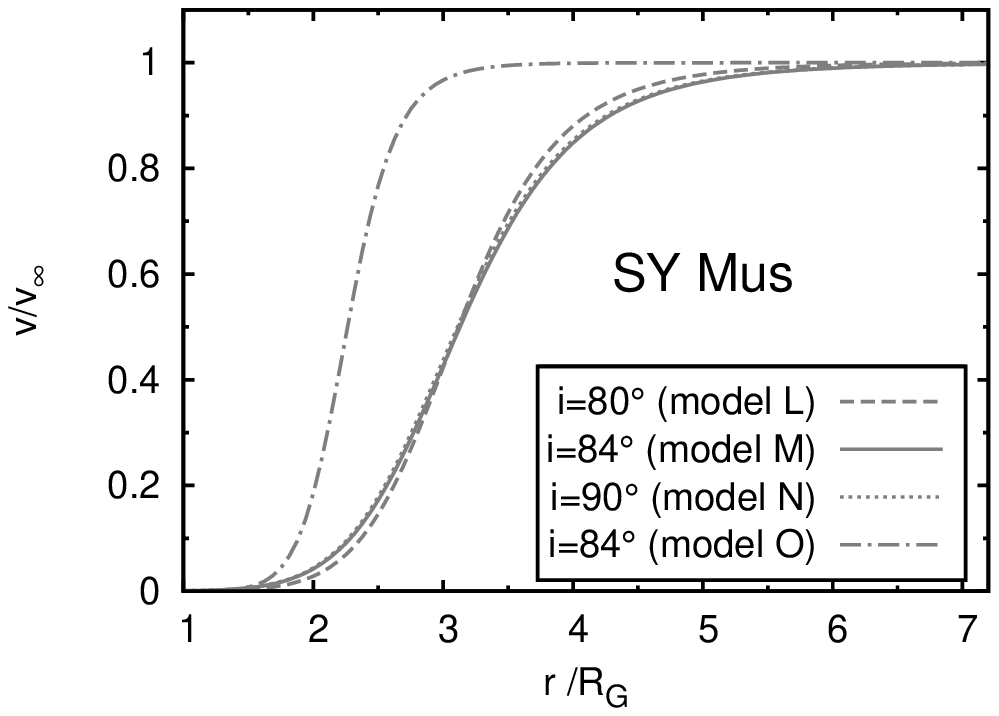}}
\resizebox{\hsize}{!}{\includegraphics[angle=0]{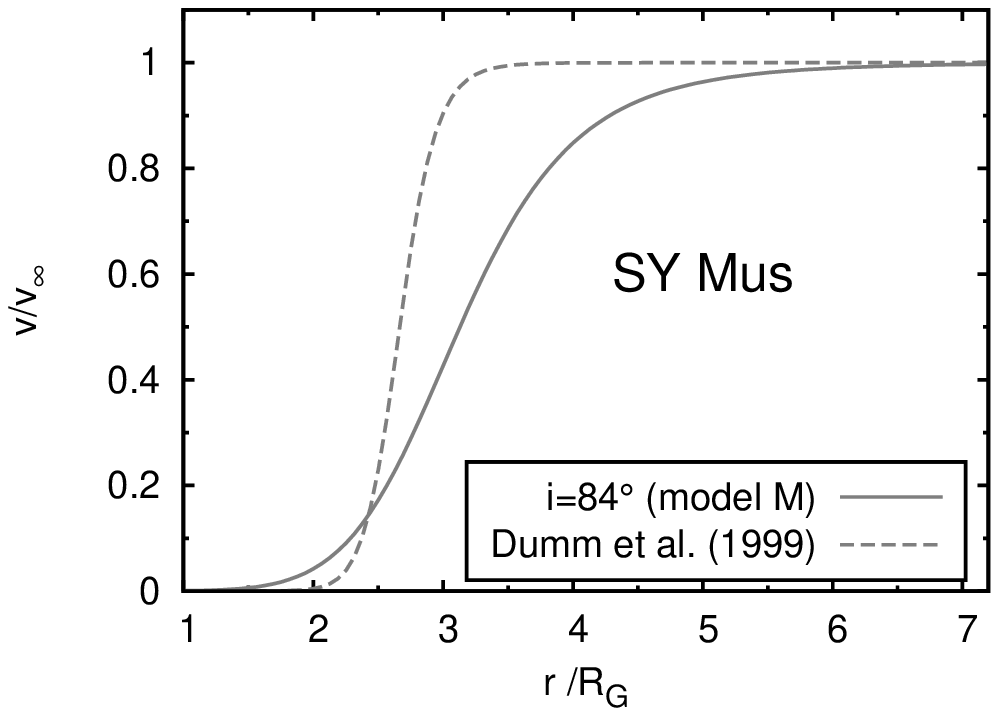}}
\end{center}
\caption[]{
WVPs of our $n_{\rm H^0}(b)$ models 
(Table~\ref{vysl}). 
The bottom panel compares our model for SY~Mus with that of 
\cite{du99}, which does not calculate ionization of the wind 
(Sect.~\ref{s:effect}). 
          }
\label{windFig}
\end{figure}
%
%
\begin{table}[t!]
\caption{Spherical equivalent of the mass-loss rate, 
          $\dot{M}_{\rm sp}$ (Eq.~(\ref{eq:dotm})), 
          corresponding to model parameters in Table~\ref{vysl} 
          for $v_\infty = 30$\kms.}
\label{tabMdot}
\centering
\begin{tabular}{lccccc}
\hline\hline
Object & i    & $\dot{M}_{\rm sp}$\,[\myr]  & model  \\
\hline
EG~And 
&$70^{\circ}$ & $2.11\times 10^{-6}$ & I \\
&$80^{\circ}$ & $1.80\times 10^{-6}$ & J \\
&$90^{\circ}$ & $1.58\times 10^{-6}$ & K \\
SY~Mus
&$80^{\circ}$ & $4.26\times 10^{-6}$ & L \\
&$84^{\circ}$ & $3.30\times 10^{-6}$ & M \\
&$90^{\circ}$ & $3.24\times 10^{-6}$ & N \\
&$84^{\circ}$ & $1.30\times 10^{-6}$ & O \\
\hline
\end{tabular}
\end{table}
%
%
\section{Discussion}
\label{dis}

\subsection{$v_\infty$ and $\dot{M}$ for giants in S-type 
            symbiotic stars}
\label{vnekdotM}

According to \cite{du86}, terminal velocities of winds from cool 
giants are $<100$\kms\ with decreasing values for later spectral 
types. 
For EG~And, \cite{cr06} determined $v_\infty = 75$\kms\ from 
the radial velocity of the absorption component of the 
$\rm Mg^+$ resonance doublet measured on the spectrum 
during the inferior conjunction of the giant. 
From molecular absorption band analysis, \cite{du99} referred 
$v_\infty$  to be in the range of 10--30\kms for M giant winds. 
From radial velocities of the \ha\ absorption component, 
\cite{sk+01} derived $v_\infty = 32\pm 6$\kms\ for 
a \ion{M4.5}{iii} giant in AX~Per. Values of 
$v_\infty = 20 - 30$\kms\ can be inferred from the position 
of the violet shifted component of the Raman scattered 
O$^{5+}$ 1032\,\AA\ line \citep[][]{schmid+99}.

Mass-loss rates from giants in symbiotic binaries are usually 
determined in the context of the STB model, in which 
a fraction of the spherically symmetric wind from the giant 
is ionized by the hot component. Then, measuring the wind 
nebular emission allows us to determine the corresponding 
mass-loss rate. 
In the radio, \cite{sea+93} measured flux density at 3.6\,cm 
for a sample of 99 symbiotic stars. Using the \cite{wb75} 
relation, corrected for the shape of the neutral part of 
the wind as given by the parameter $X^{\rm H+}$ in the STB model, 
they derived $\dot M \ga 10^{-7}$\myr\ for RGs in S-type 
symbiotic stars. A similar conclusion was reached by \cite{mi02} 
based on analysis of radio emission at 1.3\,mm. 
In the UV/optical/near-IR, we can calculate the model emission 
measure, which is determined by parameters of the giant's wind. 
Here we use a technique of disentangling the composite continuum of 
symbiotic binaries to extract the nebular component 
from the observed spectrum. In this way, \cite{sk05} determined that
$\dot M$ for giants in 15 well-observed, S-type symbiotic stars is  a few $\times 10^{-7}$\myr. 
%
Another possibility is provided by the parameter $X^{\rm H+}$, 
whose definition (see Eq.~(\ref{XH+})) allows us to determine 
$\dot M$ for a given set of binary parameters. 
In this way, \cite{mu91} estimated typical values of 
$\dot M \approx 10^{-7}$\myr\ for the RGs in 
their sample of 12 S-type symbiotic stars. 

With respect to our approach in this paper, it is important 
to stress that these values of $\dot M$ represent the total 
mass-loss rates, which are independent 
of the line of sight \citep[see STB;][]{wb75}. 

\subsection{Effect of ionization to WVP and $\dot M$}
\label{s:effect}

Since both \cite{du99} and this work applied the same method to the same 
data to determine the WVP of the giant in SY~Mus; we can test 
the effect of the wind ionization by comparing 
the corresponding velocity profiles. 

The bottom panel of Fig.~\ref{windFig} compares the WVP of the 
giant in SY~Mus derived from the column density model of 
\cite{du99}, which does not calculate ionization of the wind, 
and from our ${n}_{\rm H^0}(b)$ model (Eq.~(\ref{nHnum})), which 
includes the effect of the wind ionization.  
For $r \la 2\,R_{\rm G}$, both models require a low $v(r)$, 
which reflects a high column density on the lines of sight 
passing in the vicinity of the cool giant. 
However, for $r \ga 2\,R_{\rm G}$ the wind of the model by Dumm et al. 
 accelerates much faster than our model. 
This is given by a different particle density $N_{\rm H}(r)$
along the line of sight in the ${n}_{\rm H}(b)$ model 
(Eq.~(\ref{nHr})) and our ${n}_{\rm H^0}(b)$ 
model (Eq.~(\ref{nHnum})) constrained by the same value of 
${n}_{\rm H^0}^{\rm obs}(b)$. 
This is because in the former case, we integrate the column 
density of total hydrogen, i.e. from the observer to $\infty$, 
while in the latter case, we integrate only throughout the neutral 
wind, i.e. the integration ends at the intersection 
with the $\rm H^0$/$\rm H^+$ boundary. 
Since $v(r)\propto 1/N_{\rm H}(r)$, 
a lower $N_{\rm H}(r)$ in the former case implies a higher 
$v(r)$, which is the case of the Dumm's et al. wind model. 
At $r \ga 5\,R_{\rm G}$, $v(r)\rightarrow v(\infty)$ in 
both models. 

Similarly, ionization of the giant's wind also affects 
the corresponding model mass-loss rate. A larger 
$N_{\rm H}(r)$ along the line of sight in our 
models with respect to models without ionization corresponds 
to a larger $\dot M$ because ${\dot M} \propto N_{\rm H}(r)$. 
For example, our 
$\dot M_{\rm sp} = 3\times 10^{-6}$\myr\ for SY~Mus (Table~3) 
is a factor of $\sim 6$ larger than the Dumm et al.~(1999) 
value of $\sim 5\times 10^{-7}$\myr. Accordingly, we can conclude that the influence of the wind 
ionization in determining the WVP and $\dot M$ is fairly 
significant. 

\subsection{Focusing of the wind}
\label{sect:focusing}

Since both EG~And and SY~Mus are high-inclination systems, our 
line of sight pointing to their WDs always passes through the 
near-orbital-plane region and, thus, the column densities and 
WVP, as determined by our method, reflect characteristics 
of this area. 
We used the continuity equation (\ref{cont}) to derive the column density models given 
by Eqs.~(\ref{nHr}) and (\ref{nHnum}). However, the $n_{\rm H^0}(b)$ models correspond to 
the spherical equivalent of the mass-loss rate, 
$\dot{M}_{\rm sp} \ga 10^{-6}$\myr
(Table~\ref{tabMdot}), which is a factor $\ga 10$ higher than the total mass-loss rate 
from giants in S-type symbiotic stars (see Sect.~\ref{vnekdotM}). 
%
As the column densities were measured at the near-orbital-plane region (eclipsing binaries) and the corresponding 
$\dot{M}_{\rm sp} >> \dot{M}_{\rm total}$, the wind from the giant has to be focused 
towards the orbital plane.

This result is consistent with a 3D hydrodynamic simulations of 
the circumstellar mass distribution in the symbiotic binary RS Oph, which show a mass enhancement in the orbital plane as compared 
to the polewards directions \citep[see][]{walder+08}. 
  Also, focusing of the wind from normal giants in S-type symbiotic 
  binaries towards the orbital plane can be caused by their slow 
  rotation, as suggested by \cite{sc15}. 
However, simplifying assumptions in the model by \cite{sc15}
cannot describe  the column density distribution 
of the circumstellar material around the giant in detail in the same way as we measure 
in the real case (see Appendix A). 
%

\section{Conclusion}
\label{conc}

We modelled hydrogen column densities $n_{\rm H^0}^{\rm obs}$ 
of neutral wind from RGs in eclipsing S-type symbiotic stars 
EG~And and SY~Mus with the aim at determining the WVP and the 
corresponding $\dot M$. 
We determined  WVPs of giants in the form of Eq.~(\ref{wind}), 
obtained by inversion of the Abel's integral operator for the 
hydrogen column density function (Eq.~(\ref{nH})), derived 
 under the assumption of the spherically symmetric wind 
(Eq.~(\ref{cont})). In our approach, we fitted $n_{\rm H^0}^{\rm obs}$ values 
by the function $n_{\rm H^0}(b)$ (Eq.~(\ref{nHnum})), which 
integrates the column density only within the neutral fraction 
of the giant's wind. This represents the main novelty with 
respect to previous analyses, in which the effect of the 
wind ionization was not calculated \citep[e.g.][]{du99}. 

Our approach revealed 
that the spherical equivalent of the mass-loss rate from 
giants in EG~And and SY~Mus 
$\dot{M}_{\rm sp} \ga 10^{-6}$\myr\ 
(Sect.~\ref{sect:Msp}, Table~\ref{tabMdot}), which is a factor 
of $\ga 10$ higher than total rates, $\dot{M}$, determined 
by techniques that do not depend on the line of sight 
(Sect.~\ref{vnekdotM}). This finding suggests that the wind from giants in S-type 
symbiotic stars is not spherically symmetric. Because the measured column densities correspond to those at 
the near-orbital-plane region (eclipsing binaries) and 
$\dot{M}_{\rm sp} >> \dot{M}_{\rm total}$, the wind is focused towards the binary 
orbital plane. 
%
%
As a result, the enhanced wind at the orbital plane allows the WD 
to accrete from the giant's wind more effectively than in 
the spherically symmetric case. 

\begin{acknowledgements}
The authors would like to thank the referee Ken Gayley for 
suggestions that helped to substantially improve this paper. 
This article was created by the realization of the project ITMS 
No.~26220120009, based on the supporting operational Research 
and Development Programme financed by the European Regional 
Development Fund. 
This research was supported by a grant of the Slovak Academy of 
Sciences, VEGA No. 2/0002/13. 
\end{acknowledgements}

%
%
\appendix
\section{Comparison with the wind compression model}

Recently, \cite{sc15} (Paper I) applied the wind compression 
zone model \citep[WCZ; see][]{bjorkcass93,igncassbjork96} 
to the stellar wind of slowly rotating RGs in S-type 
symbiotic binaries. 
In this way, they showed that the high wind-mass transfer 
efficiency in these systems, as required by the 
high luminosities of their hot components, can be caused 
by the wind focusing towards the orbital plane. 

Here, we calculate the ${n}_{\rm H}(b)$ and ${n}_{\rm H^0}(b)$ 
functions using the WCZ models A--H of Paper~I 
(see Table~2 therein), and compare these functions with our solution 
using Eq.~(\ref{nHr}) and Eq.~(\ref{nHnum}), respectively. 
However, to judge the comparison, one has to take into account 
that the WCZ models of Paper~I represent only a stationary 
case including just the kinematic of the giant's wind with 
a $\beta$-law velocity profile. 
On the other hand, the measured ${n}_{\rm H^0}^{\rm obs}(b)$ 
values result from the real density distribution of the 
circumstellar material around RGs, the nature of which is not 
included in the WCZ model. 
Therefore, we do not claim to fit the 
${n}_{\rm H^0}^{\rm obs}(b)$ measurements with the WCZ model, 
but present just a comparison to learn more about a possible 
origin of their differences. 

\subsection{Density distribution in the wind compression model}

The rotation of a star can lead to the compression of its wind 
towards the equatorial regions. 
The distribution of its density as a function of the radial distance 
$r$ from the RG's centre and the polar angle $\theta$ 
is expressed as 
\begin{equation}
  N_{\rm H}(r,\theta) =
        \frac{\dot{M}}{4\pi r^2 \mu m_{\rm H}v_{\rm r}(r)}
                \left(\frac{d\mu_{\rm m}}{d\mu_{0}}\right)^{-1},
\label{eq:nh}
\end{equation}
where $\dot{M}$ is the mass-loss rate from the star, 
$v_{\rm r}(r)$ is the radial component of the wind velocity, 
and the geometrical factor $d\mu_{\rm m}/d\mu_{0}$ describes the 
compression of the wind due to rotation of the star. 
A $\beta$-law WVP is considered in the model. 
Main assumptions are briefly summarized in Paper~I, in 
detail by \cite{lamcass99}, and originally formulated by 
\cite{bjorkcass93}. 

\subsection{Ionization boundaries in the WCZ model}

The $\rm H^0$/$\rm H^+$ boundary in the WCZ model is also 
obtained by the solution of the parametric equation 
(\ref{XH+=f}), but the parameter $X^{\rm H+}$ has its 
standard form, 
\begin{equation}
  X^{\rm H+} = \frac{4\pi \mu^2 m_{\rm H}^2}
              {\alpha_{\rm B}({\rm H},T_{\rm e})}
              p L_{\rm ph}\left(\frac{v_{\infty}}
              {\dot{M}}\right)^2, 
\label{eqn:X}  
\end{equation} 
with the same denotation and meaning of parameters as 
in Eq.~(\ref{XH+}), and the function 
\begin{equation}
  f(u,\vartheta) = \int_{0}^{u_{\varphi}(b)} \frac{u^2}
      {z^4
      \left(1-\frac{\tilde{b}}{z}\frac{R_{\rm G}}{p}\right)^{2\beta} 
      \left(\frac{d\mu_{\rm m}}{d\mu_0}\right)^2} {\rm d}u 
\label{eqn:calcX} 
\end{equation}
includes the geometrical factor of the wind compression and 
the $\beta$-law WVP. Distances $r$ and $l$ 
are expressed here in units of the stars separation $p$, i.e. 
$z = r/p$, $u = l/p$ ($u_{\varphi}(b) = l_{\varphi}(b)/p$), 
$z = \sqrt{u^2+1-2u\sin i \cos\vartheta}$ , and 
$\tilde{b} = 1-(a/v_{\infty})^{1/\beta}$ is the parameter 
of the wind model (see Paper~I). 

\subsection{Column densities in the WCZ model}

Using the density distribution in the WCZ model, 
$N_{\rm H}(r,\theta)$, 
the column density 
of the total hydrogen along the line of sight containing 
the WD is calculated as 
\begin{equation}
  n_{\rm H}(b) = \int_{-\infty}^{\infty}
                 N_{\rm H}(r,\theta)\,{\rm d}l,
\label{eqn:coldeninfty}
\end{equation}
which is compared to Eq.~(\ref{nHr}) of this paper. 
%
Similarly, the column density restricted to the $\rm H^0$ zone 
only is calculated as 
\begin{equation}
  n_{\rm H^0}(b) = 
            \int_{-\infty}^{l_{\varphi}(b)}
            N_{\rm H}(r,\theta)\,{\rm d}l,
\label{eqn:colden}
\end{equation}
which is compared to Eq.~(\ref{nHnum}). 

\begin{figure*}
\centering
\begin{center}
\resizebox{\hsize}{!}{\includegraphics[angle=0]{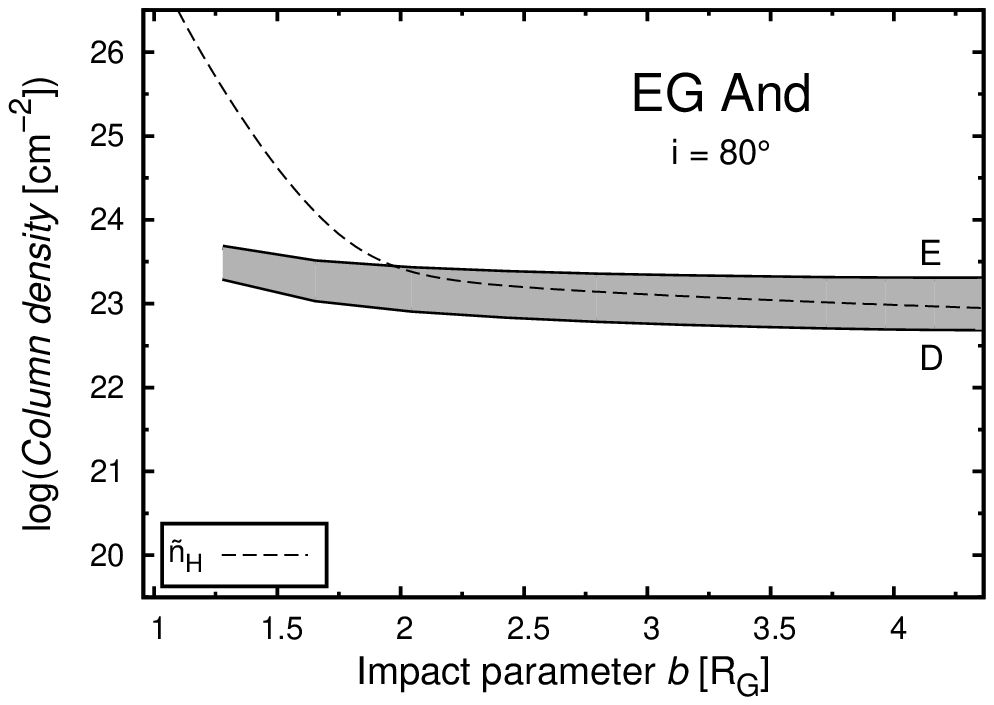}
                      \includegraphics[angle=0]{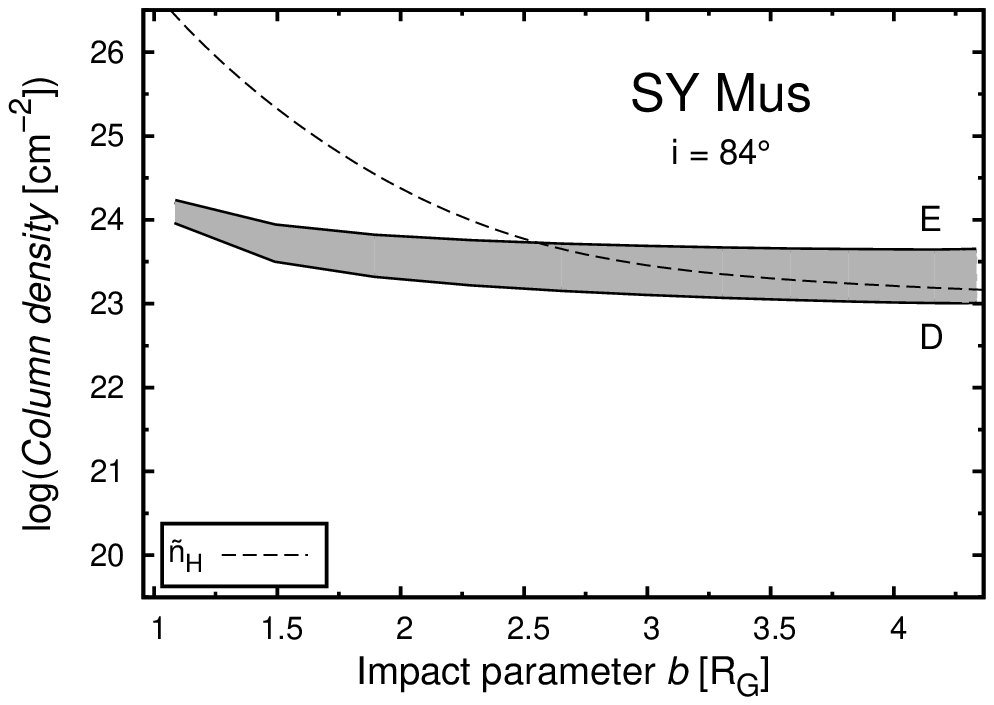}}
\resizebox{\hsize}{!}{\includegraphics[angle=0]{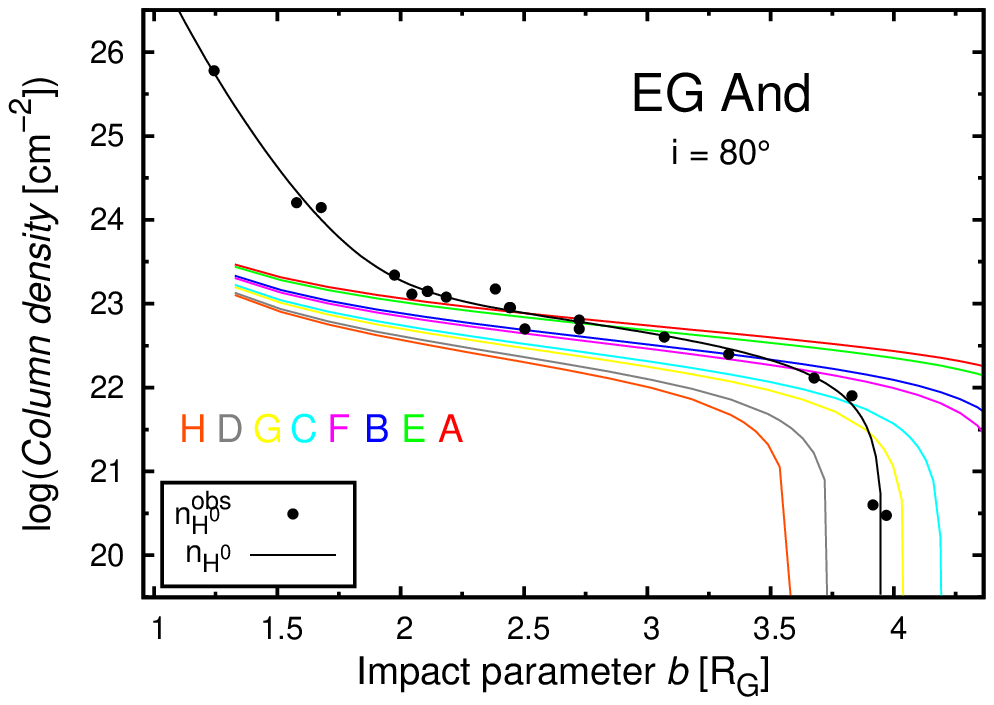}
                      \includegraphics[angle=0]{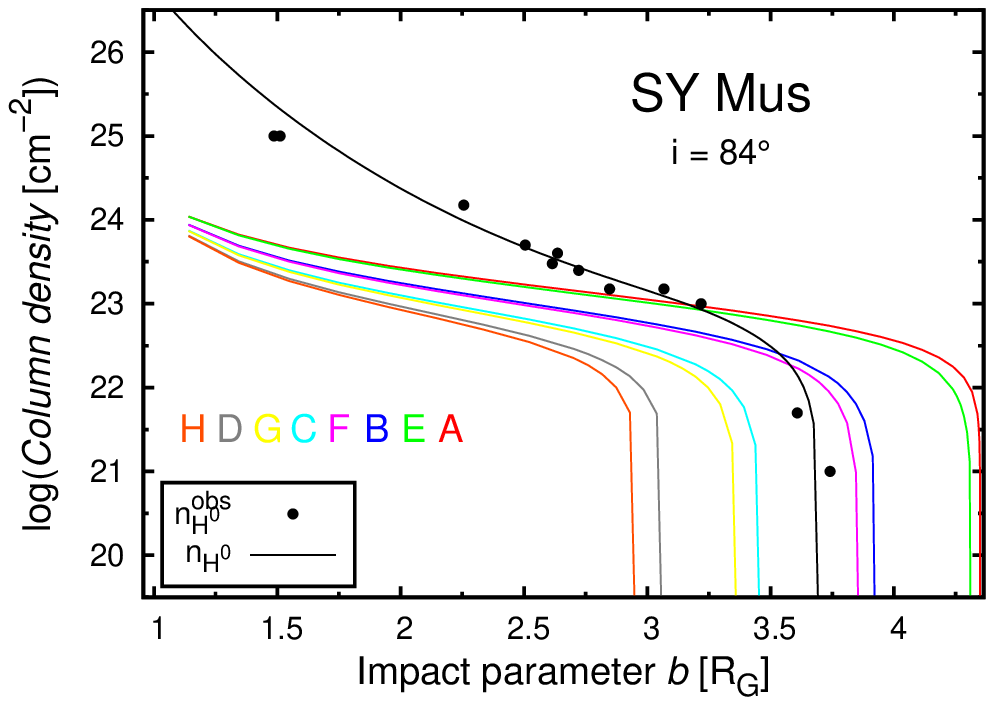}}
\end{center}
\caption[]{
Comparison of hydrogen column densities of this paper with 
those calculated according to the WCZ models of Paper~I. 
Top panels show examples of models J and M (Eq.~(\ref{nHr}); 
dashed lines) with those calculated according to 
Eq.~(\ref{eqn:coldeninfty}) for WCZ models A--H of Paper~I 
(grey area bounded by models E and D). 
Bottom panels compare the same models, but calculated throughout 
the neutral zone only (this paper: Eq.~(\ref{nHnum}), solid black 
line; WCZ models of Paper~I: Eq.~(\ref{eqn:colden}), coloured 
lines). 
}
\label{fig:comp}
\end{figure*}

Figure~\ref{fig:comp} provides this comparison for models 
J and M of this paper and WCZ models of Paper~I using the input 
parameters in Table~\ref{t:par}. 
%
%
For $b \la 2$, there is a large difference in all cases. 
This difference can be caused by the large uncertainties in determining of 
${n}_{\rm H^0}^{\rm obs}$ values from low-resolution spectra
and by the presence of other absorbing effects (e.g. bound-free 
transitions and the Rayleigh attenuation in the continuum) 
that are not included in the model (\ref{rayleighflux}). 
   For further distances ($b \ga 2-2.5$) and the column density of 
the total hydrogen, $n_{\rm H}(b)$ functions in the WCZ models 
(\ref{eqn:coldeninfty}) are comparable with $\tilde{n}_{\rm H}(b)$ 
(top panels of Fig.~\ref{fig:comp}). 
However, in models including the effect of ionization, 
$n_{\rm H^0}(b)$ functions in the WCZ models (\ref{eqn:colden}), 
which match the measured values within $3.5\ga b \ga 2.5$, 
bend at larger $b$ because the $\rm H^0$ zone opens more at 
the near-orbital-plane region because of the wind compression 
there (bottom panels of Fig.~\ref{fig:comp}). 
  On the other hand, the WCZ models (\ref{eqn:colden}) with 
the bend around $b = 3.5 - 4$, are below the 
$n_{\rm H^0}^{\rm obs}(b)$ values\ because these models 
(H, D, G, C for EG~And and C, F, B for SY~Mus, see the figure) 
correspond to a smaller compression, and thus to lower 
$N_{\rm H}(r,\theta)$ densities implying lower 
$n_{\rm H^0}(b)$ functions (\ref{eqn:colden}). 

\begin{figure}
\centering
\begin{center}
%
\resizebox{6.5cm}{!}{\includegraphics[angle=0]{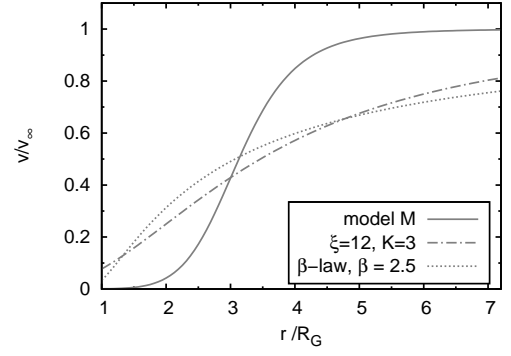}}
\end{center}
\caption[]{
Examples of the $v(r)$ profiles 
and the $\beta$-law wind of Paper~I. 
          }
\label{windXiK}
\end{figure}

Here we note that Paper~I suggests the WCZ model just as 
a possible efficient wind-mass transfer mode in S-type symbiotic 
binaries. However, simplifying assumptions of the model do not 
allow us to provide any information on the real structure of 
the wind from the rotating giant. As a result, corresponding 
$n_{\rm H^0}(b)$ functions (\ref{eqn:colden}) do not match 
$n_{\rm H^0}^{\rm obs}(b)$ values, which are given by the 
real density distribution in the binary. 
Nevertheless, the amount of the total hydrogen at the plane of 
observations in the WCZ models (\ref{eqn:coldeninfty}) matches 
that derived from the column density measurements 
(top panels of Fig.~\ref{fig:comp}). 
Finally, a good fit of the $n_{\rm H^0}^{\rm obs}(b)$ quantities 
by the function (\ref{nHnum}) is due to a very flexible WVP 
(\ref{wind}), whose variables do not have clear 
physical meaning, however. Example $v(r),$ which matches the $\beta$-law wind of 
Paper~I, is shown in Fig.~\ref{windXiK}. 
%
\begin{table*}
\caption{Input parameters for WCZ models of Paper~I}
\label{t:par}
\centering
\begin{tabular}{cccccccccc}
\hline
\hline
Object  & $R_{\rm G}$ [\ro] & $p$ [\ro]$^{1)}$ & $\dot{M}$ [\mo] &
$T_{\rm h}$ [K] & $L_{\rm h}$ [\lo]$^{2)}$ & $a$ [\kms] & $\beta$ &
$v_{\infty}$ [\kms] & $v_{\rm rot}$ [\kms]$^{3)}$ \\
\hline
 EG~And & 75 & 330 & 2$\times 10^{-7}$ &  95\,000 &   77 
        & 1  & 2.5 & 20 - 50           & 6 - 9.5 \\
 SY~Mus & 86 & 376 & 4$\times 10^{-7}$ & 110\,000 & 1200 
        & 1  & 2.5 & 20 - 50           & 6 - 9.5 \\
\hline
\multicolumn{10}{l}{
{\bf Notes:} 
$^{1)}$ parameters $R_{\rm G}$ and $p$ as in Sect.~4.4, 
$^{2)}$ parameters $\dot{M}$, $T_{\rm h}$ and $L_{\rm h}$ 
        are from \cite{sk05}, 
                   } \\
\multicolumn{10}{l}{
~~~~~~~~~~~~$^{3)}$ parameters of the wind, $a$, $\beta$, $v_{\infty}$,
        and $v_{\rm rot}$ are adopted from Paper~I. 
                   } \\
\end{tabular}
\end{table*}
%
%
\end{document}